\documentclass[3p,times]{elsarticle}

 \usepackage{ecrc}

\volume{00}

\firstpage{1}

\journalname{Nuclear Physics A}

\runauth{J. Uphoff et al.}

\jid{nupha}

\jnltitlelogo{Nuclear Physics A}

\usepackage{amssymb}

\usepackage[
   centertags, 
   sumlimits,  
   intlimits,  
   namelimits, 
]{amsmath} %
\usepackage{amssymb}

\biboptions{numbers,square,sort&compress}

\usepackage{overpic}

\begin{document}

\begin{frontmatter}




\title{Open heavy flavor and J/$\psi$ at RHIC and LHC}


\author[label1]{Jan Uphoff}
\author[label1]{Oliver Fochler}
\author[label2]{Zhe Xu}
\author[label1]{Carsten Greiner}

\address[label1]{Institut f\"ur Theoretische Physik, Johann Wolfgang 
Goethe-Universit\"at Frankfurt, Max-von-Laue-Str. 1, 
D-60438 Frankfurt am Main, Germany}
\address[label2]{Department of Physics, Tsinghua University, Beijing 100084, China}

\begin{abstract}
The suppression and collective flow of open and hidden heavy flavor in ultra-relativistic heavy ion collisions is studied with the partonic transport model \emph{Boltzmann Approach to MultiParton Scatterings} (BAMPS). Charm and bottom quarks interact with the rest of the medium via elastic scatterings with an improved Debye screening and a running coupling. Heavy flavor electron data from RHIC can only be described if the cross section is multiplied with $K=3.5$ to mimic the contribution from radiative processes. With the same value for $K$  we compare to elliptic flow data of $D$ mesons at the LHC and find a good agreement. However, the nuclear modification factor of $D$ mesons, non-prompt $J/\psi$, muons, and electrons at the LHC  is underestimated in our calculations. In addition, we present first BAMPS results on the  suppression and elliptic flow of $J/\psi$ mesons, which are treated in the same framework.
\end{abstract}


\end{frontmatter}


\section{Introduction}

Experimental measurements at the Relativistic Heavy Ion Collider (RHIC) \cite{Adams:2005dq,Adcox:2004mh} and Large Hadron Collider (LHC) \cite{Muller:2012zq} reveal that in ultra-relativistic heavy ion collision a hot and dense medium is created which exhibits exciting properties like collective flow and jet quenching. This medium consists of deconfined quarks and gluons and, hence, is called quark gluon plasma (QGP). Heavy quarks in particular are an interesting probe to study the properties of the medium since they can only be produced in initial hard parton scatterings or in the early phase of the medium evolution due to their large mass \cite{Uphoff:2010sh}. Therefore, they traverse the medium right from the beginning for a rather long time, interact with other medium particles, participate in the flow and lose energy.

Due to flavor conservation the number of heavy quarks stays constant during hadronization. Open heavy flavor mesons can then be measured (directly or indirectly), which reveals important information about the heavy quark distribution and hence the properties of the QGP. At the LHC, ALICE has reconstructed $D$ mesons directly and measured their nuclear modification factor $R_{AA}$ \cite{Abelev:2012nj} and elliptic flow $v_2$ \cite{alice_hp_d_v2}. In addition, also the $B$ meson $R_{AA}$ has been measured via non-prompt $J/\psi$ by the CMS collaboration \cite{Chatrchyan:2012np}. More indirect are the measurements of heavy flavor muons \cite{Abelev:2012qh} and electrons \cite{Adare:2010de,Pachmayer:2011wq} which have contributions from both $D$ and $B$ mesons.
All measurements show a large energy loss of heavy quarks in the QGP and a strong participation in the collective flow.

Complimentary to open heavy flavor mesons with only one heavy quark are hidden heavy flavor mesons like $J/\psi$ which consist of a charm and an anti-charm quark. Lattice calculations indicate that $J/\psi$ states can survive in the QGP to some extent \cite{Mocsy:2007jz}. The most interesting observable here is the nuclear modification factor of $J/\psi$. However, in contrast to the open heavy flavor $R_{AA}$ it is not dominated by energy loss, but by the total number of $J/\psi$ in the medium. Nevertheless, this observable is also strongly influenced by cold nuclear matter effects like shadowing which makes it a challenging task for theoretical explanations.


\section{The partonic transport model BAMPS}
\emph{Boltzmann Approach to MultiParton Scatterings} (BAMPS) \cite{Xu:2004mz,Xu:2007aa} is a partonic transport model which solves the Boltzmann equation for on-shell particles and pQCD interactions. For light partons all $2\rightarrow 2$ as well as $2 \leftrightarrow 3$ processes are implemented. On the heavy flavor sector so far only binary scatterings of heavy quarks with light partons are active, namely, 
$g Q \rightarrow g Q  $ and 
       $ q Q \rightarrow q Q$.
The divergent $t$ channel of those processes is regularized with a screening mass $\mu$ which is determined by matching energy loss calculations with the Born cross section to results from hard thermal loop calculations \cite{Gossiaux:2008jv,Peshier:2008bg,Uphoff:2011ad}. The comparison of both results shows that the screening mass $\mu$ is of the order of the Debye mass $m_D$, $\mu^2 = \kappa m_D^2$ with $\kappa = 1/(2e) \approx 0.2$. Furthermore, the running of the coupling is explicitly taken into account. More details on both can be found in Ref.~\cite{Uphoff:2011ad}.
We note that currently for light partons neither a running coupling nor an improved Debye screening is employed, which we plan, however, to do in the future.
Inelastic heavy flavor processes \cite{Abir:2011jb} are currently being studied in BAMPS and results will be presented soon. 

In addition, $J/\psi$ mesons have been added to BAMPS which can be dissociated by a gluon, $J/\psi +g \rightarrow c + \bar c$ \cite{Bhanot:1979vb}, and regenerated via the back reaction of this process. Furthermore, cold nuclear matter effects, that is in BAMPS shadowing, nuclear absorption, and the Cronin effect, are introduced, which is outlined in detail in Ref.~\cite{Uphoff:2011fu}.

\section{Results on open heavy flavor}

The running coupling and the improved screening procedure, which reproduces the energy loss from hard thermal loop calculations, effectively enhances the heavy quark cross section with the medium. Quantitative comparisons \cite{Uphoff:2010bv,Uphoff:2011ad,Fochler:2011en,Meistrenko:2012ju,Uphoff:2012gb} show that elastic processes are responsible for a large fraction of the energy loss of heavy quarks. However, they are not able to reproduce the data of the nuclear modification factor or elliptic flow of any heavy flavor particle species. This is not too surprising since we expect that radiative $2 \rightarrow 3$ processes also play an important role and that both processes together should account for the measured suppression and flow. The quantitative contribution of radiative processes will be studied in an forthcoming investigation. 
In this paper we mimic the radiative influence by effectively increasing the elastic cross section by a factor $K=3.5$ which fits the heavy flavor elliptic flow data from RHIC \cite{Uphoff:2012gb}. Simultaneously, the $R_{AA}$ data at RHIC can also be described with the same parameter. Whether this effective description is valid and radiative contributions really boil down to simply multiplying the elastic cross section by a constant factor will be studied with BAMPS in the future.

\begin{figure}[t]
\begin{minipage}[t]{0.49\textwidth}
\centering
\includegraphics[width=1.0\textwidth]{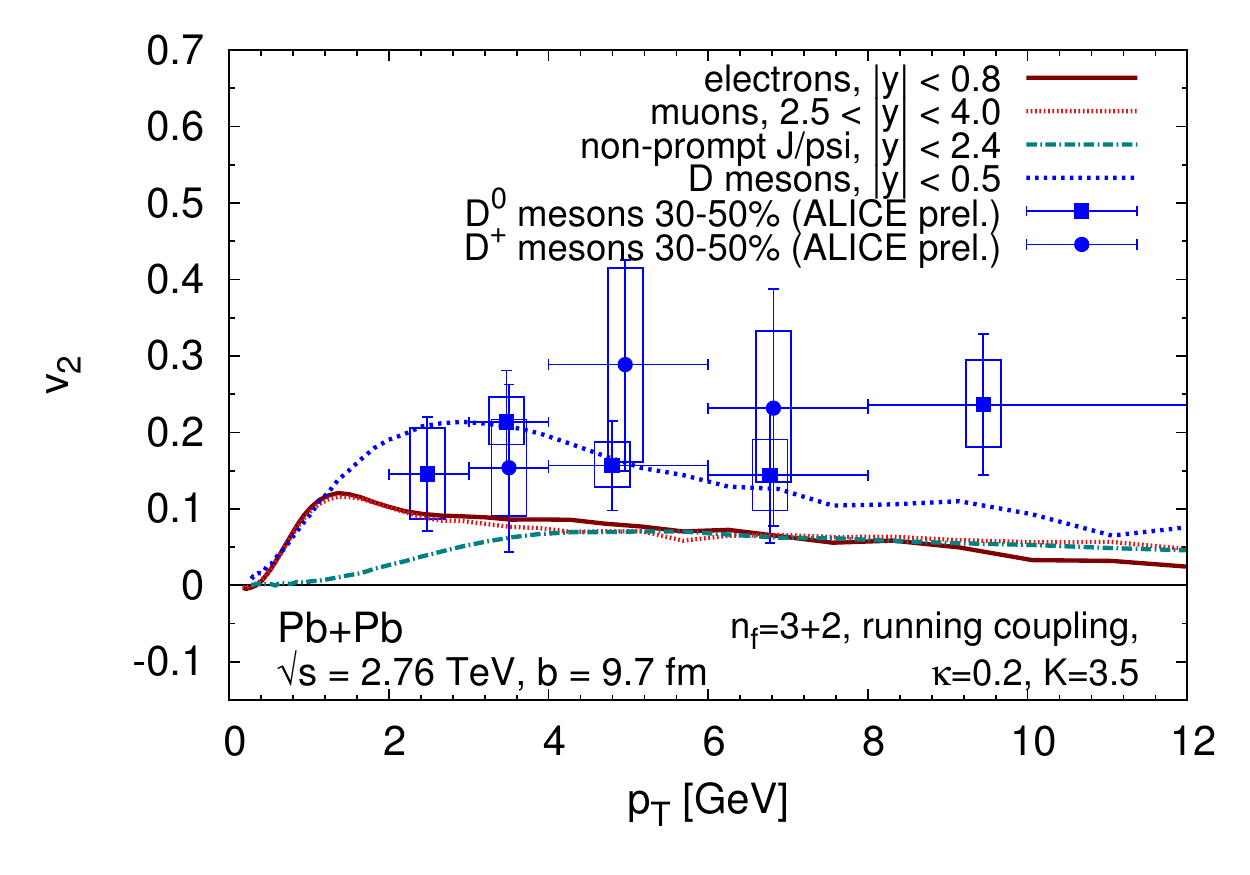}
\end{minipage}
\hfill
\begin{minipage}[t]{0.49\textwidth}
\centering
\includegraphics[width=1.0\textwidth]{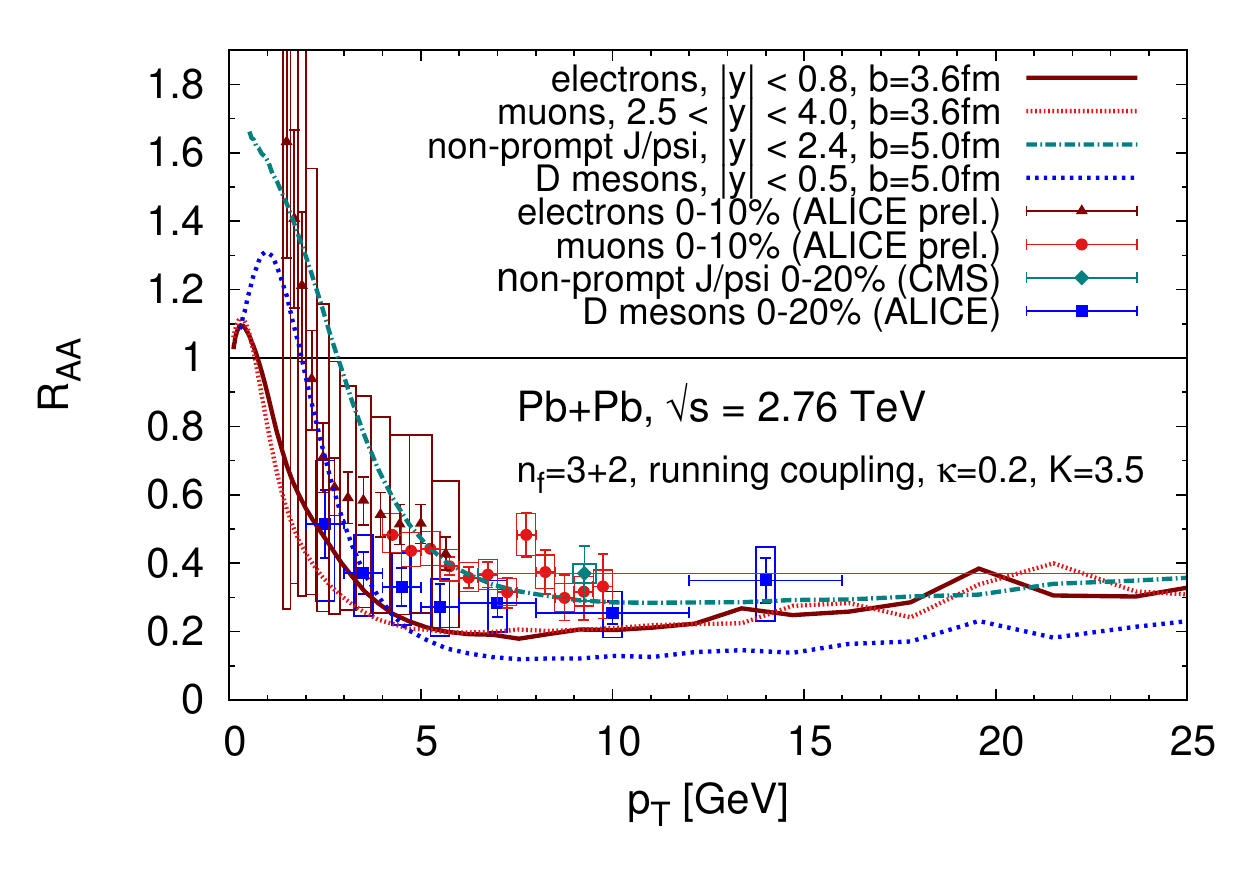}
\end{minipage}
\caption{Elliptic flow $v_2$ (left) and nuclear modification factor $R_{AA}$ (right) of $D$ mesons, non-prompt $J/\psi$, muons, and electrons as a function of transverse momentum for Pb+Pb collisions at LHC together with data \cite{alice_hp_d_v2,Abelev:2012nj,Pachmayer:2011wq,Masciocchi:2011fu}. As a note, the centrality classes of the $R_{AA}$ data are not always the same. The impact parameter of the BAMPS calculations is matched to the corresponding centrality class to be able to compare to the data.
}
\label{fig:v2_raa_overview_lhc}
\end{figure}

Figure~\ref{fig:v2_raa_overview_lhc} gives an overview of the BAMPS results on the elliptic flow and nuclear modification factor of all heavy flavor particles which can be measured at the LHC.
The experimental data is also plotted wherever it is available. The BAMPS predictions for the elliptic flow of $D$ mesons are in good agreement with the data points over the whole transverse momentum range. This hints that the effective binary cross section with $K=3.5$, which fits the RHIC data, can also describe the elliptic flow at LHC, although the experimental errors are still too large to draw any definite conclusion.
The $v_2$ of $D$ mesons is considerably larger than that of non-prompt $J/\psi$ due to the mass difference of charm and bottom quarks. The elliptic flow of electrons and muons stemming from both $D$ and $B$ mesons is somewhere in between.

The nuclear modification factor of all heavy flavor species is at the lower edge of the error bars or even below it. Therefore, the suppression in BAMPS with $K=3.5$ is too strong, although it agrees with the RHIC $R_{AA}$ of heavy flavor electrons. However, at RHIC the error bars especially at high transverse momentum are rather large. Therefore it could be possible that also at RHIC scaled elastic collisions alone cannot describe both $v_2$ and $R_{AA}$ simultaneously and radiative collisions must be taken into account.

\section{Suppression and elliptic flow of $J/\psi$}
\begin{figure}
\begin{minipage}[t]{0.49\textwidth}
\centering
\begin{overpic}[width=1.0\textwidth]{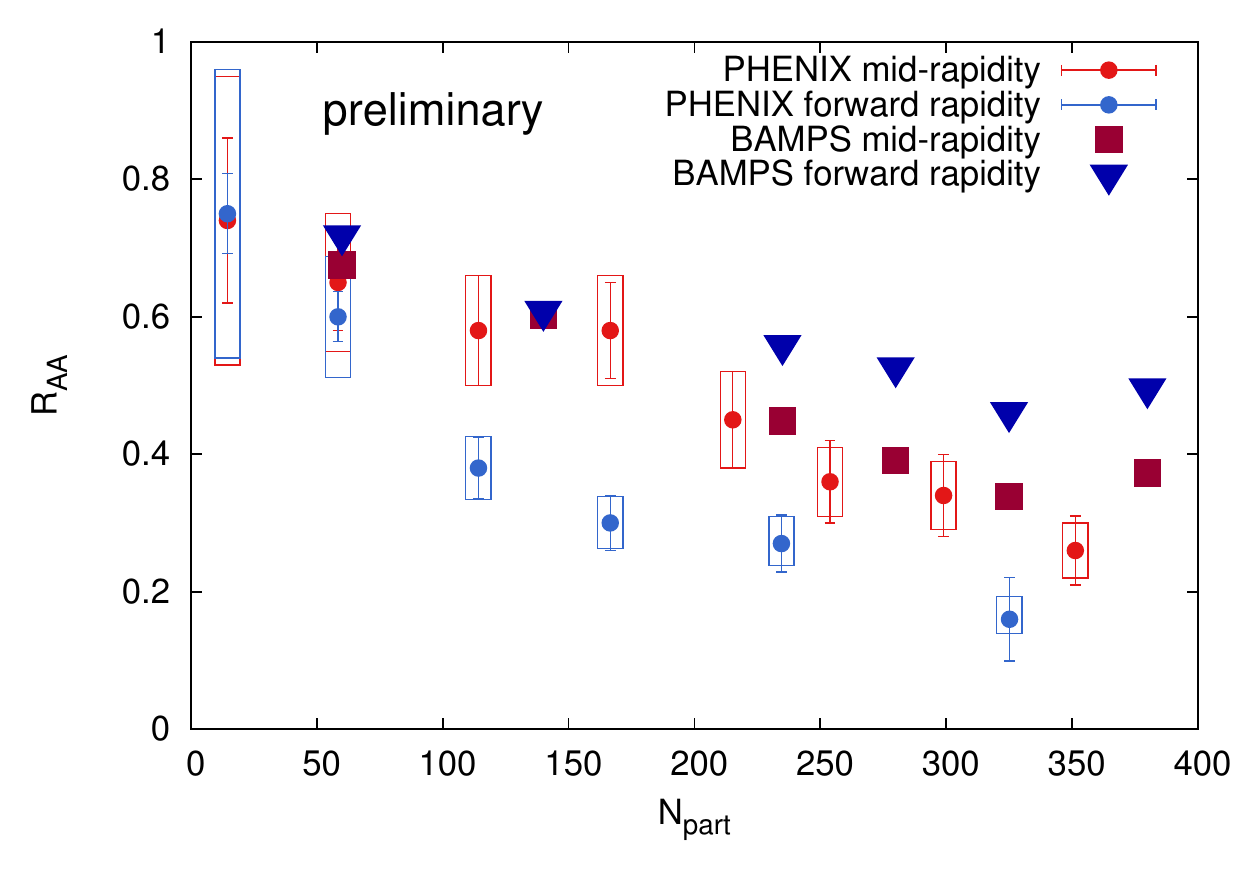}
\put(23,18){\footnotesize $\tau_0 = 0.6 \,{\rm fm}$} 
\end{overpic}
\end{minipage}
\hfill
\begin{minipage}[t]{0.49\textwidth}
\centering
\begin{overpic}[width=1.0\textwidth]{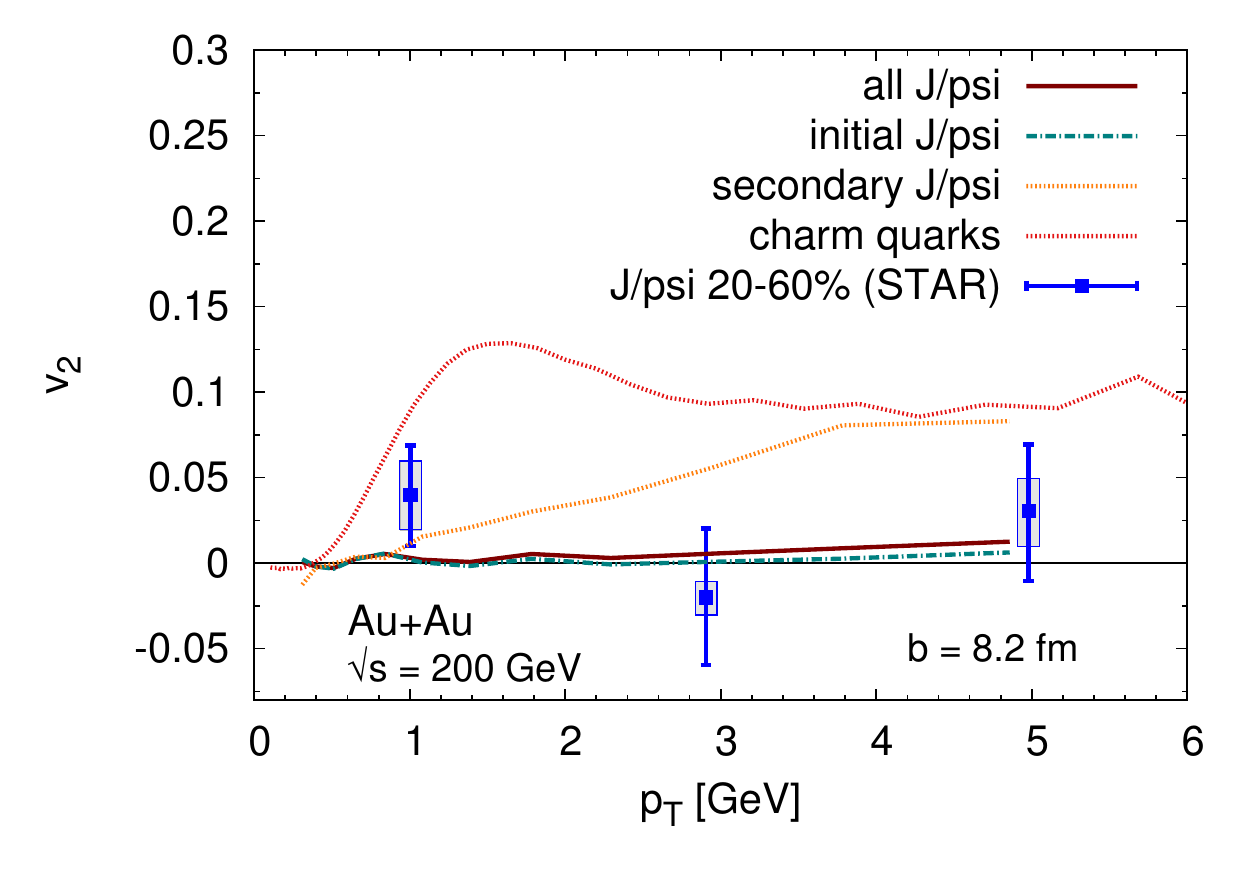}
\put(28,55){\footnotesize preliminary} 
\end{overpic}
\end{minipage}
\caption{Left panel: $R_{AA}$ of $J/\psi$ at mid-rapidity $|y| < 0.35$ and forward rapidity $1.2 < y < 2.2$ for Au+Au collisions at RHIC as a function of the number of participants, together with experimental data \cite{Adare:2006ns}. Right panel: Elliptic flow of $J/\psi$ with data \cite{Powell:2011gf}. For comparison the charm quark $v_2$ is also shown.}
\label{fig:jpsi_raa_rhic_npart}
\end{figure}

First BAMPS results on the suppression of $J/\psi$ mesons at RHIC were presented in Ref.~\cite{Uphoff:2011fu}. 
In Fig.~\ref{fig:jpsi_raa_rhic_npart} the $R_{AA}$ of $J/\psi$ at RHIC is depicted as a function of the number of participants.
For this calculation  a formation time of $\tau_0 = 0.6 \,{\rm fm}$ is employed for $J/\psi$ mesons to prevent early melting when the temperature cannot be properly defined.
The experimental data  of the $R_{AA}$ at mid-rapidity is well described. However, our results at forward rapidity underestimate the suppression for central and semi-central events.

In this section we want to focus primarily on the elliptic flow of $J/\psi$. Experimental measurements at RHIC showed that the $v_2$ of $J/\psi$ is very small \cite{Powell:2011gf}. This is in contradiction with the regeneration picture where the flow of the charm quarks should be transferred to the $J/\psi$. BAMPS is an ideal framework to study this in more detail since it reproduces the $D$ meson flow (cf. previous section) and also allows recombination of charm quarks to $J/\psi$.
The right panel of Fig.~\ref{fig:jpsi_raa_rhic_npart} shows that in BAMPS even with regeneration the elliptic flow of all $J/\psi$ is compatible with the 
data.

\begin{figure}[t]
\begin{minipage}[t]{0.49\textwidth}
\centering
\includegraphics[width=1.0\textwidth]{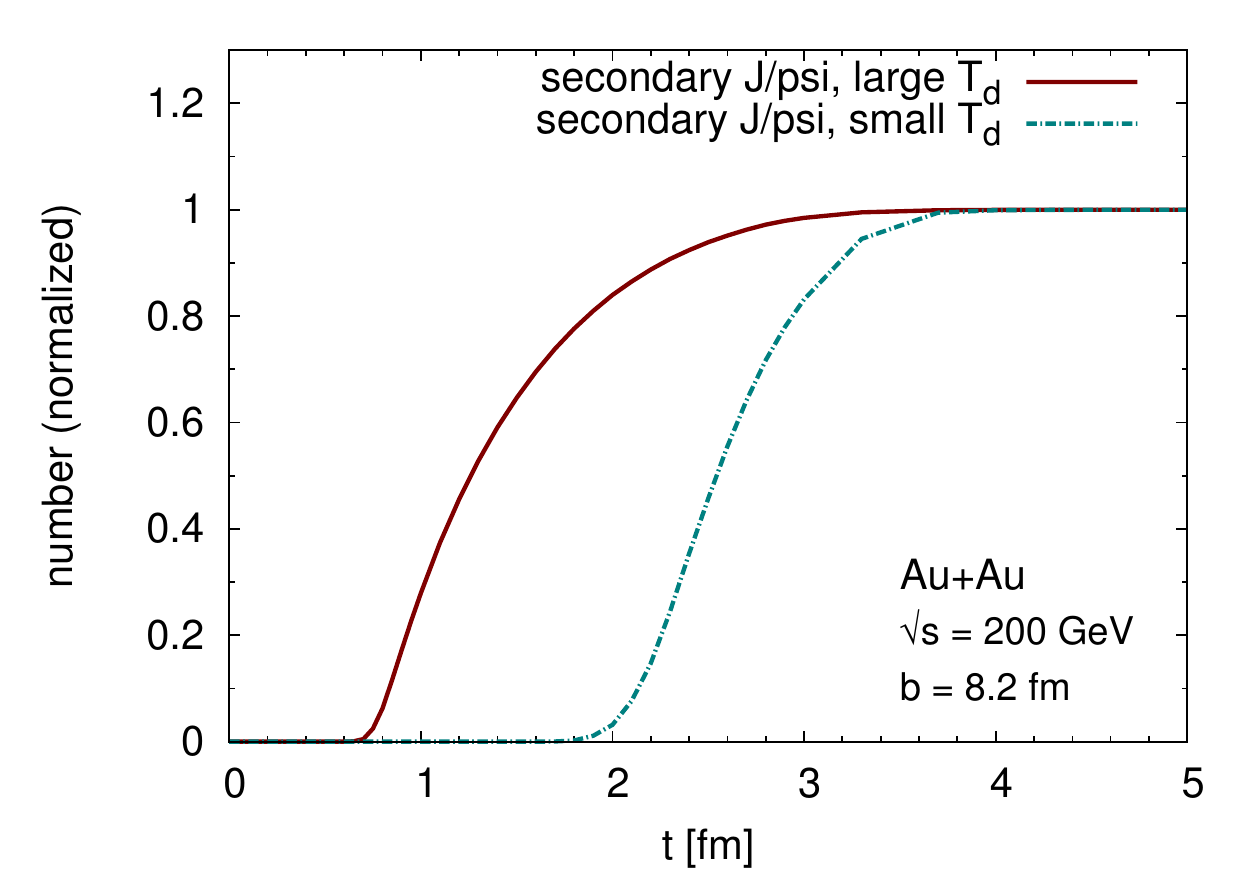}
\end{minipage}
\hfill
\begin{minipage}[t]{0.49\textwidth}
\centering
\includegraphics[width=1.0\textwidth]{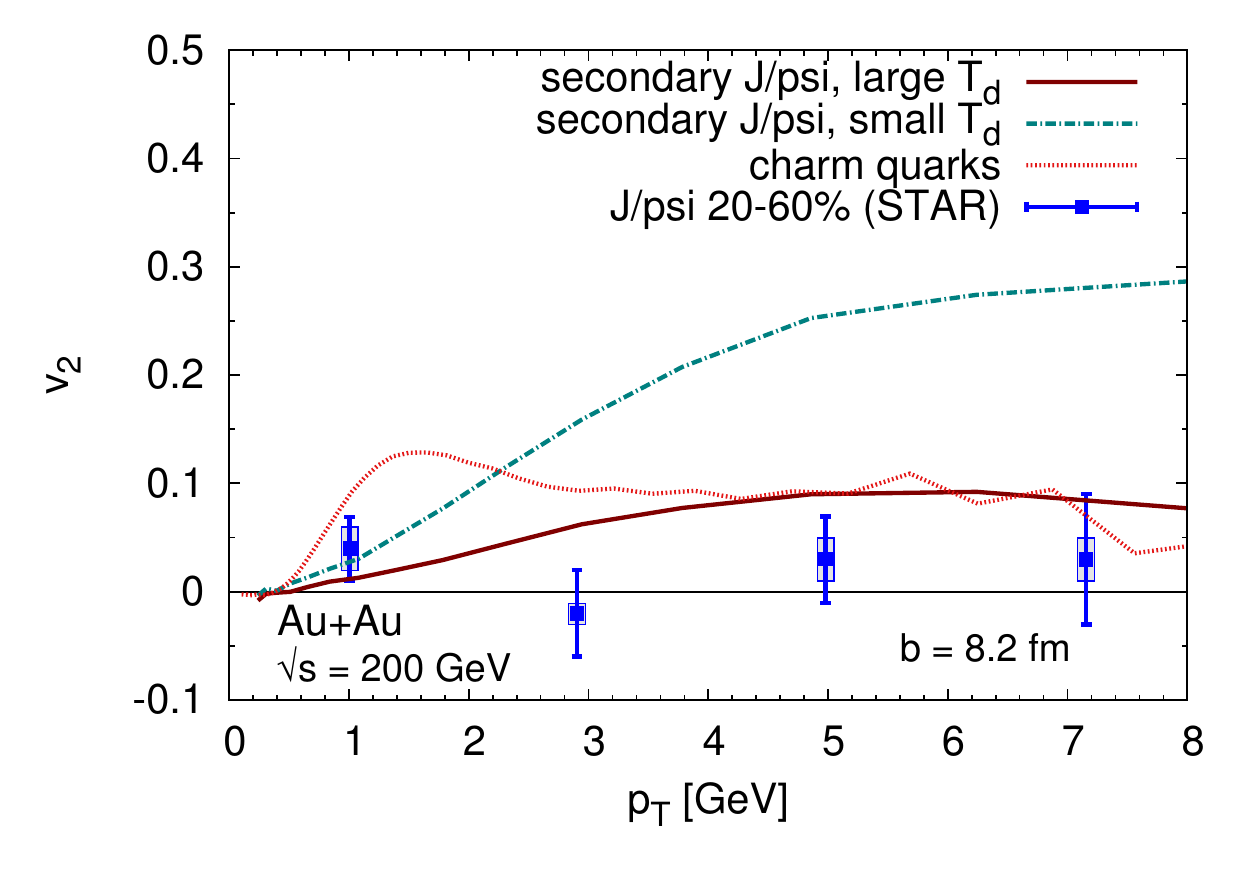}
\end{minipage}
\caption{Time evolution of the number of secondary $J/\psi$ (normalized) and their elliptic flow as a function of transverse momentum at the end of the QGP phase at RHIC for small and large $J/\psi$ melting temperature $T_d$. For comparison, the charm $v_2$ and experimental data on $J/\psi$ \cite{Powell:2011gf} are also shown. As a note, the contribution of initial $J/\psi$ with vanishing $v_2$ is not depicted.
}
\label{fig:v2_jpsi}
\end{figure}

To estimate the effect of regeneration on flow in the following qualitative study, we consider only the elliptic flow of secondary $J/\psi$, which are created in the medium from the recombination of two charm quarks. Initial $J/\psi$ has (nearly) vanishing $v_2$ and must be taken into account if one wants to compare to data.
In Fig.~\ref{fig:v2_jpsi} we show the number of secondary $J/\psi$ as a function of time and their elliptic flow as a function of transverse momentum for two scenarios. 
In one scenario the $J/\psi$ melting temperature is large. Hence, secondary $J/\psi$ can be formed already in the early phase of the medium evolution where the temperature is still large. In the other scenario we choose a small melting temperature which only allows secondary $J/\psi$ production at late times. 

$J/\psi$ from the first scenario only have a small $v_2$ since they are produced early and their constituent charm quarks could not build up a sizeable flow. The second scenario corresponds more to the coalescence picture. The $J/\psi$ $v_2$ is shifted to twice as large values at twice as large $p_T$ compared to the charm flow, $v_2^{J/\psi} (p_T) \approx 2 v_2^c(p_T/2)$. Here the fraction of initial $J/\psi$ at large $p_T$ must be considerably larger than secondary $J/\psi$ to be compatible with the data.

\section{Conclusions}
We presented recent BAMPS results on the elliptic flow and nuclear modification factor of several heavy flavor particles, namely $D$ mesons, non-prompt $J/\psi$, muons, and electrons. Using only elastic collisions of heavy quarks with medium particles with an improved Debye screening inspired by hard thermal loop calculations and a running coupling the RHIC heavy flavor electron data can only be reproduced with scaling the binary cross section with $K=3.5$. With the same parameter we find a good agreement with the $D$ meson elliptic flow data at LHC. However, the $R_{AA}$ of $D$ mesons, non-prompt $J/\psi$, and muons are underestimated. Furthermore, we showed some preliminary results on the suppression and elliptic flow of  $J/\psi$ with a focus on flow of secondary $J/\psi$ that stem from charm quarks which develop flow within the same framework.





\bibliography{hq}

\begin{thebibliography}{10}

\bibitem{Adams:2005dq}
STAR, J.~Adams {\em et~al.},
\newblock Nucl. Phys. {\bf A757}, 102 (2005), nucl-ex/0501009.

\bibitem{Adcox:2004mh}
PHENIX, K.~Adcox {\em et~al.},
\newblock Nucl. Phys. {\bf A757}, 184 (2005), nucl-ex/0410003.

\bibitem{Muller:2012zq}
B.~M\"uller, J.~Schukraft, and B.~Wyslouch,
\newblock (2012), 1202.3233.

\bibitem{Uphoff:2010sh}
J.~Uphoff, O.~Fochler, Z.~Xu, and C.~Greiner,
\newblock Phys. Rev. {\bf C82}, 044906 (2010), 1003.4200.

\bibitem{Abelev:2012nj}
ALICE Collaboration, B.~Abelev {\em et~al.},
\newblock (2012), 1203.2160.

\bibitem{alice_hp_d_v2}
ALICE Collaboration, G.~Ortona,
\newblock (2012),
\newblock these proceedings.

\bibitem{Chatrchyan:2012np}
CMS Collaboration, S.~Chatrchyan {\em et~al.},
\newblock (2012), 1201.5069.

\bibitem{Abelev:2012qh}
ALICE Collaboration, B.~Abelev {\em et~al.},
\newblock (2012), 1205.6443.

\bibitem{Adare:2010de}
PHENIX, A.~Adare {\em et~al.},
\newblock Phys.Rev. {\bf C84}, 044905 (2011), 1005.1627.

\bibitem{Pachmayer:2011wq}
ALICE Collaboration, Y.~Pachmayer,
\newblock J.Phys.G {\bf G38}, 124186 (2011), 1106.6188.

\bibitem{Mocsy:2007jz}
A.~Mocsy and P.~Petreczky,
\newblock Phys. Rev. Lett. {\bf 99}, 211602 (2007), 0706.2183.

\bibitem{Xu:2004mz}
Z.~Xu and C.~Greiner,
\newblock Phys. Rev. {\bf C71}, 064901 (2005), hep-ph/0406278.

\bibitem{Xu:2007aa}
Z.~Xu and C.~Greiner,
\newblock Phys. Rev. {\bf C76}, 024911 (2007), hep-ph/0703233.

\bibitem{Gossiaux:2008jv}
P.~B. Gossiaux and J.~Aichelin,
\newblock Phys. Rev. {\bf C78}, 014904 (2008), 0802.2525.

\bibitem{Peshier:2008bg}
A.~Peshier,
\newblock (2008), 0801.0595.

\bibitem{Uphoff:2011ad}
J.~Uphoff, O.~Fochler, Z.~Xu, and C.~Greiner,
\newblock Phys.Rev. {\bf C84}, 024908 (2011), 1104.2295.

\bibitem{Abir:2011jb}
R.~Abir, C.~Greiner, M.~Martinez, M.~G. Mustafa, and J.~Uphoff,
\newblock Phys.Rev. {\bf D85}, 054012 (2012), 1109.5539.

\bibitem{Bhanot:1979vb}
G.~Bhanot and M.~E. Peskin,
\newblock Nucl. Phys. {\bf B156}, 391 (1979).

\bibitem{Uphoff:2011fu}
J.~Uphoff, K.~Zhou, O.~Fochler, Z.~Xu, and C.~Greiner,
\newblock PoS {\bf BORMIO2011}, 032 (2011), 1104.2437.

\bibitem{Uphoff:2010bv}
J.~Uphoff, O.~Fochler, Z.~Xu, and C.~Greiner,
\newblock Nucl.Phys. {\bf A855}, 444 (2011), 1011.6183.

\bibitem{Fochler:2011en}
O.~Fochler, J.~Uphoff, Z.~Xu, and C.~Greiner,
\newblock J.Phys.G {\bf G38}, 124152 (2011), 1107.0130.

\bibitem{Meistrenko:2012ju}
A.~Meistrenko, A.~Peshier, J.~Uphoff, and C.~Greiner,
\newblock (2012), 1204.2397.

\bibitem{Uphoff:2012gb}
J.~Uphoff, O.~Fochler, Z.~Xu, and C.~Greiner,
\newblock (2012), 1205.4945.

\bibitem{Masciocchi:2011fu}
ALICE, S.~Masciocchi,
\newblock J.Phys.G {\bf G38}, 124069 (2011), 1109.6436.

\bibitem{Adare:2006ns}
PHENIX, A.~Adare {\em et~al.},
\newblock Phys. Rev. Lett. {\bf 98}, 232301 (2007), nucl-ex/0611020.

\bibitem{Powell:2011gf}
STAR Collaboration, C.~B. Powell,
\newblock (2011), 1111.6944.

\end{thebibliography}







\end{document}